\documentclass[11pt]{iopart}
\usepackage{graphicx}
\DeclareGraphicsExtensions{.eps}
\usepackage[english]{babel}
\usepackage{amsfonts}
\usepackage{amssymb}
\expandafter\let\csname equation*\endcsname\relax
\expandafter\let\csname endequation*\endcsname\relax 
\usepackage{amsmath}
\usepackage[ansinew]{inputenc}
\usepackage{multirow}
\usepackage{mathtools}
\usepackage{wasysym}
\usepackage{color}
\usepackage{verbatim}
\usepackage{mcite}

\begin{document}

\title{Elastic depinning transition of vortex lattices in two dimensions}

\author{N Di Scala$^1$, E Olive$^1$, Y Lansac$^1$, Y Fily$^2$ and JC Soret$^1$}
\address{$^1$ GREMAN, UMR 7347, Universit\'e F. Rabelais, Parc de Grandmont, 37200 Tours, France}
\address{$^2$ Department of Physics, Syracuse University, Syracuse, New York 13244, USA}

\begin{abstract}
Large scale numerical simulations are used to study the elastic dynamics of two-dimensional vortex lattices driven on a disordered medium in the case of weak disorder. We investigate the so-called elastic depinning transition by decreasing the driving force from the elastic dynamical regime to the state pinned by the quenched disorder.
Similarly to the plastic depinning transition, we find results compatible with a second order phase transition, although both depinning transitions are very different from many viewpoints. We evaluate three critical exponents of the elastic depinning transition. $\beta = 0.29 \pm 0.03$ is found for the velocity exponent at zero temperature, and from the velocity-temperature curves we extract the critical exponent $\delta^{-1} = 0.28 \pm 0.05$. Furthermore, in contrast with charge-density waves, a finite-size scaling analysis suggests the existence of a unique diverging length at the depinning threshold with an exponent $\nu= 1.04 \pm 0.04$, which controls the critical force distribution, the finite-size crossover force distribution and the intrinsic correlation length. Finally, a scaling relation is found between velocity and temperature with the $\beta$ and $\delta$ critical exponents  both independent with regard to pinning strength and disorder realizations.
\end{abstract}

\maketitle

\section{Introduction}
\label{intro}

The physics of disordered elastic systems is relevant for numerous systems, {\it e.g.} 
interfaces like magnetic \cite{Lemerle1998} or ferroelectric \cite{Paruch2005} domain walls, contact lines \cite{Moulinet2004}, crack propagation \cite{Bouchaud2002}, or periodic structures like charge-density waves \cite{Gruner1988,*Brazovskii2004} (CDWs), vortex lattices in type II superconductors \cite{Blatter1994,*Nattermann2011,*Giamarchi2002}, colloids \cite{Pertsinidis2008} or Wigner crystals \cite{Giamarchi2003}. It is essential to understand the response of these systems to an external driving force, such as current-induced Lorentz force for vortices, electric field for charge-density waves, or magnetic field for domain walls. The competition between elasticity and disorder in all these systems leads to a great variety of phases. In particular the system remains pinned up to a critical driving force $F_c$ and starts sliding above $F_c$. The transition from a pinned state to a sliding one is known as the \emph{depinning transition}. It has been suggested by Fisher \cite{Fisher1985}, in the context of CDWs, that the depinning transition could be regarded as a standard equilibrium critical phenomenon where the reduced force $f=(F-F_c)/F_c$ and the velocity $v$ would respectively act as the control parameter and the order parameter. This idea has proven useful in many other systems. This analogy suggests the existence of critical exponents, in particular $\beta$ which describes the power-law response of the velocity at zero temperature 
$v \sim f^{\beta}$.
However the analogy with standard critical phenomena is limited since the depinning transition is by nature a non-equilibrium transition. For example, unlike equilibrium phase transitions no divergent steady-state correlation length scale exists for an elastic line when approaching the depinning from below \cite{Kolton2006,*Kolton2009}. Connections between the depinning transition of elastic interfaces in random media and non-equilibrium phase transitions into absorbing states are reported in the literature \cite{Hinrichsen2000,*Alava2002,*Odor2004,*Bonachela2007}. The pinned phase where the dynamics is frozen may be considered as an absorbing state in which the system may enters but cannot leave. 
Two universality classes have been proposed for the interface growth. The linear growth of the interface is described in the quenched Edwards-Wilkinson equation within the random field Ising class \cite{Nattermann1992,Narayan1993,Kim2001}, while Kardar-Parisi-Zhang nonlinearities in this equation make the depinning transition closely related to directed percolation \cite{Tang1992,*Makse1998}. Non-equilibrium phase transition into absorbing states has also been shown in periodic systems with quenched disorder like colloids: when periodically sheared a transition was observed from a diffusive liquid to a pinned state \cite{Corte2008}, while in the case of driven colloids the transition separates plastic flow from a pinned state \cite{Reichhardt2009}. The critical exponents found in both studies are consistent with aborbing state transitions in the universality classes of directed percolation or conserved directed percolation.

\indent When the strength of the underlying disorder is strong, the theoretical description of the phenomenon is difficult and the question about the nature of the depinning transition remains an open problem. In periodic systems with a 2D displacement field of dimension ($N=2$) (e.g., superconductor vortices, colloids and Wigner crystals), simulations \cite{Jensen1988b,*Ryu1996,*Faleski1996,*Gronbech1996,*Olson1998a,*Kolton1999,*Fangohr2001,*Reichhardt2001,*Olive2009,Olive2006a,*Olive2008,Reichhardt2002} have shown that this situation leads to dislocations and plasticity: at the depinning threshold 
 tearing of the lattice appears where
regions of pinned particles coexist with particles flowing around them.
When the pinning of the underlying substrate is weak the statics and dynamics are dominated by the system elasticity which favors order. After the work of Fisher, further analytical and numerical works have refined the analogy with equilibrium critical phenomena to define universality classes and critical exponents (see e.g. \cite{Bustingorry2010} and references therein).
However in numerical simulations of periodic systems in $d=2$ various exponents have been found. $\beta \approx 2/3$ is found for CDWs \cite{Myers1993}, Wigner crystals \cite{Piacente2005}, and colloids \cite{Reichhardt2002}, while other works found $\beta \approx 0.5$ \cite{Pertsinidis2001} and $\beta=0.92 \pm 0.01$ \cite{Chen2004} in colloids, and $\beta=0.35$ \cite{OlsonReichhardt2011} in stripe systems. 
For superconductor vortices in $d=3$ the value $\beta=0.65$ is found \cite{Luo2007}, while in $d=2$ a surprising exponent $\beta>1$ has been reported \cite{Cao2000}.\\

\indent In this paper, we show a detailed study of the critical behavior of superconductor vortices above the elastic depinning transition, including a finite-size scaling analysis and the determination of three critical exponents. 
The elastic depinning transition investigated in this paper is very different from the plastic depinning transition studied in a previous paper \cite{Fily2010}. In the plastic case tearing of the lattice at the depinning threshold is observed with moving vortices flowing along changing interconnected channels around pinned regions of vortices. On the contrary, at the elastic depinning threshold, all vortices depin together and flow along well-defined rough static channels in a topologically ordered structure. Furthermore, the velocity is periodic in time at the elastic threshold, while it is shown to be chaotic at the plastic threshold \cite{Olive2006a,*Olive2008} with large broad-band noise at low frequency. The curvature of the velocity-force curve is also very different since the elastic depinning is caracterized by a velocity exponent $\beta <1$ whereas $\beta >1$ for the plastic depinning.

We perform large scale molecular dynamics simulations of 2D superconductor vortex lattices driven over weak random disorder. 
Our model belongs to the category of 2D periodic systems with $N=2$ and short-range interactions. It could model 3D superconductors (either conventional or layered) in an effective 2D regime, i.e., when the vortex line tension is high enough for the lines to remain straight. 
The numerical model is developed in section \ref{NM}. The depinning transition at zero temperature is detailed in section \ref{T=0} for different lattice sizes and disorder realizations. Above the depinning threshold a power law response $v_{F>F_c} \sim f^{\beta}$ is found with $\beta = 0.29 \pm 0.03$, and the transition seems to be governed by the correlation length divergence with an exponent $\nu = 1.04 \pm 0.04$. The same exponent governs the finite-size effects. 
In section \ref{T!=0} we study the depinning transition at finite temperature. A power law response for the velocity $v_{F=F_c} \sim T^{1/\delta}$ is found and the critical exponent $\delta^{-1} = 0.28 \pm 0.05$ is determined. A scaling relation for the driving force and the temperature is established. Varying system size, disorder realization and pinning strength, the scaling function and both $\beta$ and $\delta$ critical exponents show some degree of universality.
The critical exponent values of our study are discussed in section 5.

\section{Numerical model}
\label{NM}

We study $N_v$ Abrikosov vortices interacting with $N_p$ random pins in the $(x,y)$ plane. Periodic boundary conditions are used in both $x$ and $y$ directions. We consider the London limit $\lambda_L\gg\xi_s$, where $\lambda_L$ is the penetration length and $\xi_s$ is the superconducting coherence length, \emph{i.e.} we treat vortices as point particles. As in \cite{DiScalaJPCS}, the LAMMPS 
classical molecular dynamics code \cite{LAMMPS} is used to integrate the newtonian equations of motion with a velocity Verlet algorithm \cite{Allen1987}. 
We use this parallel code in order to simulate large system sizes, which is crucial to study the elastic depinning since the Larkin domains are larger than in the case of plastic dynamics. However the LAMMPS code includes an inertial term proportionnal to the particle mass. The existence of a vortex mass has been discussed in the literature \cite{Suhl1965,*Sonin1998,*Kopnin1998,*Chudnovsky2003,*Han2005,Golubchik2012}. In general vortices are considered as massless objects although the vortex mass may be significant in the clean limit. In recent experiments \cite{Golubchik2012}, it is shown that $Nb$ vortices close to $T_c \approx 9K$ have an inertial mass which is in the intermediate range between the dirty and clean limit, showing that the inertial vortex mass should be a meaningful concept.

We describe the time evolution of a vortex $i$ at a position ${\bf r}_i$ with the equation
\begin{equation}
\begin{split}
m \frac{d{^2\bf r}_i}{dt^2}+\eta \frac{d{\bf r}_i}{dt}=- & {\sum_{j \neq i}}\nabla_i U^{vv}(r_{ij})-{\sum_{p}}\nabla_i U^{vp}(r_{ip})  \\
  & +{\bf F}^L+{\bf F}_i^{\text{th}}(t)
\end{split}
\end{equation}
\noindent
where $r_{ij}$ is the distance between the vortices $i$ and $j$ located at ${\bf r}_i$ and ${\bf r}_j$, $r_{ip}$ is the distance between the vortex $i$ and the pinning site located at ${\bf r}_p$,
and $\nabla_i$ is the 2D gradient operator acting on ${\bf r}_i$.
${\bf F}^L=F{\bf \hat x}$ is the Lorentz driving force in the $x$ direction induced by an applied current, and $\eta$ is the viscosity coefficient. The Langevin force ${\bf F}_i^{\text{th}}$ describes the coupling with a heat bath and can be expressed by a thermal gaussian white noise with zero mean and variance
$$\langle F_{i,\lambda}^{\text{th}}(t) F_{j,\mu}^{\text{th}}(t') \rangle=2 \eta k_B T \delta_{ij} \delta_{\lambda \mu} \delta (t-t')$$
where $\lambda, \mu=x,y$ are the cartesian components of ${\bf F}_i^{\text{th}}$ and $k_B$ is the Boltzmann constant.
The vortex-vortex pairwise repulsive interaction is given by a modified Bessel function
$$U^{vv}(r_{ij})=\alpha_v K_0(r_{ij}/\lambda_L)$$
and the attractive pinning potential is given by
$$U^{vp}(r_{ip})=-{\alpha_p} e^{-(r_{ip}/R_p)^2}$$
where $R_p$ the radius of the pins, and $\alpha_v$ and $\alpha_p$ are tunable parameters.
We fix the strength of the vortex-vortex interaction by setting $\alpha_v=2.83\ 10^{-3}\lambda_L \epsilon_0$ where $\epsilon_0$ is a characteristic energy per unit length.
The number of pins is set to $N_p=N_v$ and their radius is $R_p=0.22\lambda_L$. The positions of the pins are chosen from a uniform random distribution. Each random sampling will be referred to as one "sample" or one "disorder realization". The average vortex distance is $a_0=\lambda_L$, and the vortex-vortex interaction is dealt with using a neighbor list method with a cutoff radius $r_c=6.5\ \lambda_L$. 
Since we are mainly interested in the permanent regimes we choose $\eta/m=0.1$ (where $m$ is the vortex mass) in the LAMMPS code such that the inertial term is small compared to the viscous term. Care has been taken to check that such ratio ensures that the long time behavior of the second order Newton's vortex dynamics is identical to the overdamped dynamics limit usually computed for the superconductor vortices. In particular we find identical vortex trajectories and identical values of the critical exponents using overdamped dynamics on smaller systems. 

We use a unit system in which $\eta=1$, $\lambda_L=1$, $\epsilon_0=1$ and $k_B=1$.
Depending of the relative strengths of the vortex-pin and vortex-vortex interaction, the dynamics can be either dominated by elasticity or disorder. Several values of the relative disorder strength $\alpha_p/\alpha_v$ have been investigated.
In sections \ref{v-f}, \ref{FSS} and \ref{delta} we choose the relative disorder strength $\alpha_p/\alpha_v \approx 5.10^{-3}$, weak enough to enforce elastic behavior (plasticity is found above $\alpha_p/\alpha_v \approx 0.01$). In section \ref{scaling} we investigate several pinning strengths in the elastic regime, from $\alpha_p/\alpha_v \approx 2.10^{-3}$ to $\alpha_p/\alpha_v \approx 7.10^{-3}$.
 Molecular dynamics simulations are performed for several system sizes, from $N_v=270$ up to $N_v=20000$ vortices, but keeping fixed vortex and pin densities. Moreover, various rectangular shaped basic cells of size $(L_x,L_y)$ have been investigated, from the almost square $(L_x,L_y)=(5, 6 \sqrt3/2) n \lambda_L$ with $n=3,8,20$ up to very elongated strips $(L_x,L_y)=(400, 20 \sqrt3/2) \lambda_L$. The strip geometry elongated in the driving force direction allowed the study of the critical depinning properties for large system sizes. Care has been taken to check that the basic cell anisotropy induced by the strip geometry does not alter the critical properties, and in particular the value of the velocity critical exponent $\beta$.\\

\section{Zero temperature}
\label{T=0}

\begin{figure}[!h]
\centering
\includegraphics[width=0.40\textwidth]{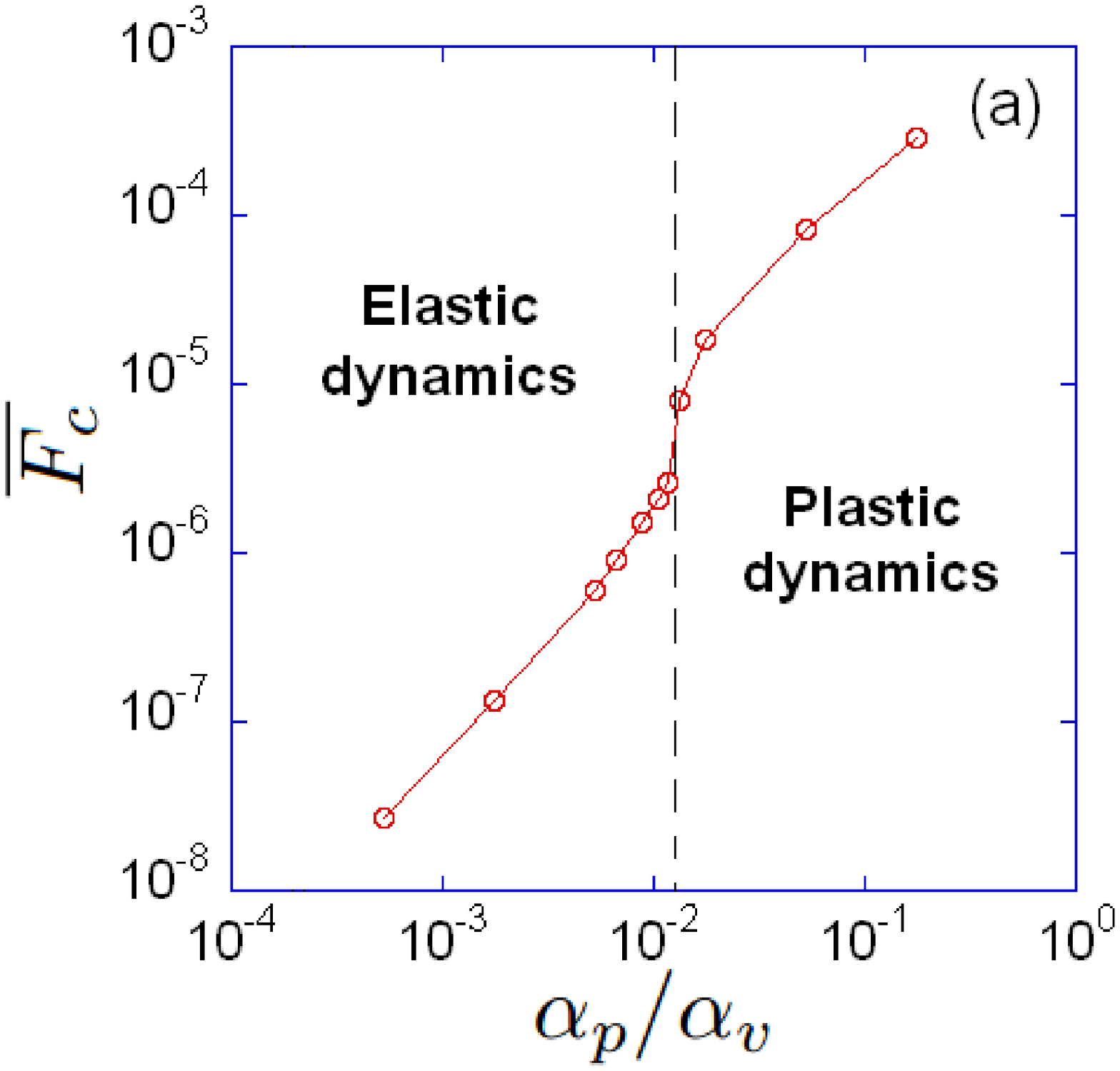}
\includegraphics[width=0.40\textwidth]{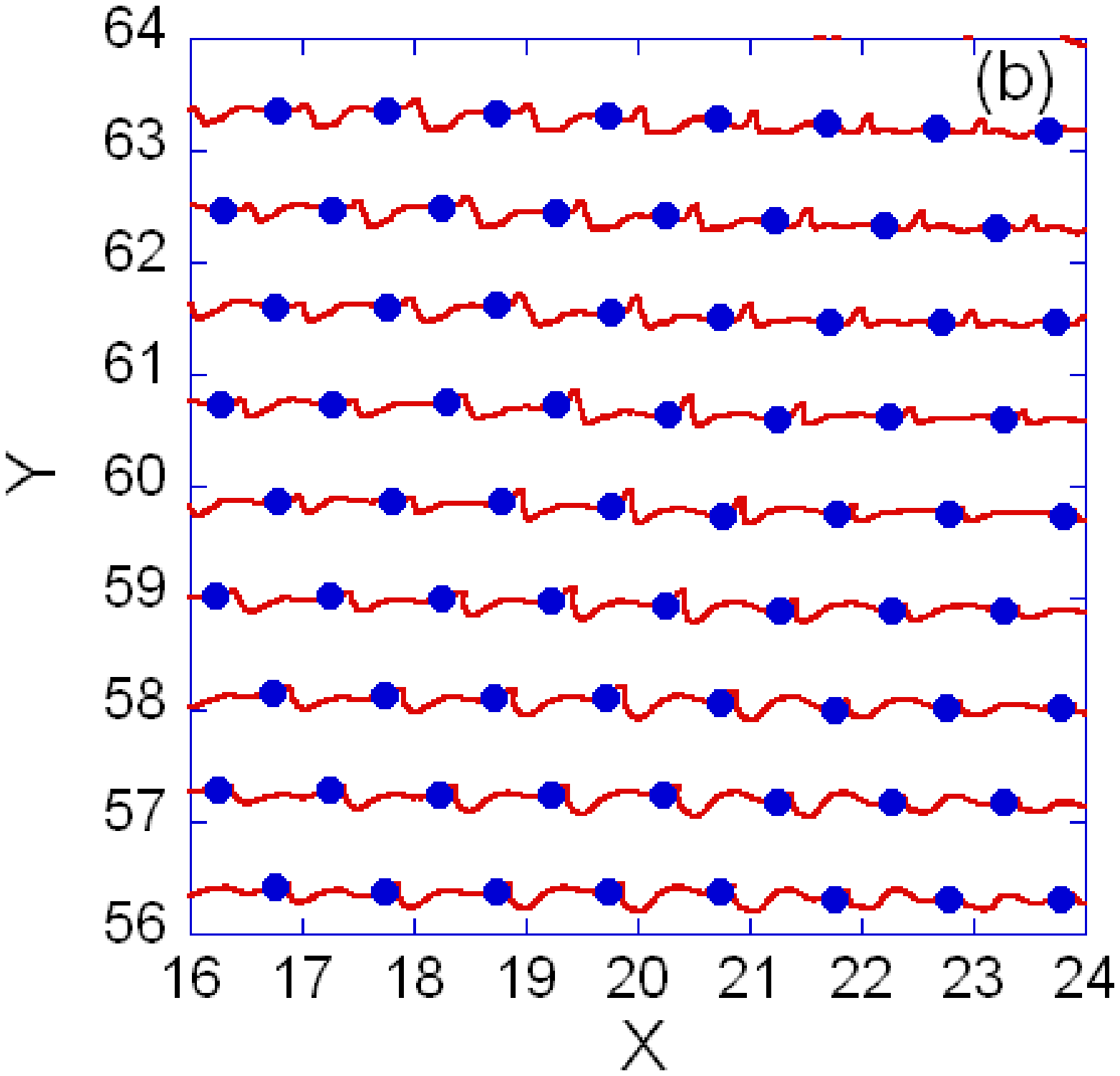}
\caption{a) Transition from elastic to plastic dynamical regimes shown by the evolution of the mean critical depinning force $\overline{F_c}$ with respect to the relative disorder strength $\alpha_p/\alpha_v$ for $N_v=8000$ vortices in a basic cell of size $(L_x,L_y)=(400, 20 \sqrt3/2) \lambda_L$. b) Typical trajectories of vortices at the elastic depinning threshold obtained for $N_v=12000$ vortices and $\alpha_p/\alpha_v \approx 5.10^{-3}$. A snapshot of the vortex positions (filled circles) at a given time is superimposed to their trajectories. For clarity only a small part of the basic cell of size $(L_x,L_y)=(100, 120 \sqrt3/2) \lambda_L$ is shown.}
\label{Fig1}
\end{figure}

For several disorder realizations we start from a perfect triangular lattice at high velocity and the driving force is slowly decreased down to the sample-dependent critical depinning force $F_c^{\text{sample}}$ below which the system is permanently pinned. We compute the mean critical depinning force $\overline{F_c}$ where the overline indicates average over disorder realizations. We show in Fig.\ref{Fig1}a the evolution of $\overline{F_c}$ with respect to the relative disorder strength $\alpha_p/\alpha_v$ for $N_v=8000$ vortices in a basic cell of size $(L_x,L_y)=(400, 20 \sqrt3/2) \lambda_L$.
The rapid increase in the mean critical depinning force indicates a crossover from elastic dynamics dominated by 
elasticity to plastic dynamics dominated by disorder: such a sharp increase was for example found in a previous work on a similar system \cite{Reichhardt2002}. In Fig.\ \ref{Fig1}b the typical trajectories of the vortices at the elastic depinning threshold are displayed for $\alpha_p/\alpha_v \approx 5.10^{-3}$. The vortices flow in  elastically coupled rough static channels and the structure is topologically ordered: all vortices depin together with the same mean velocity, which means that all vortices keep the same neighbors as they move. The dynamics is jerky and the velocity of the center of mass is periodic in time where the period corresponds to the time for each vortex to replace its preceding neighbor in the same channel.\\

\subsection{Velocity-force response}
\label{v-f}

For each system size $(L_x,L_y)$ we define the reduced velocity, $$v=\overline{<V_x^{cm}(t)>/F_c^{\text{sample}}}$$ and the reduced force $f=(F-F_c^{\text{sample}})/F_c^{\text{sample}}$ where $<V_x^{cm}(t)>$ is the time averaged longitudinal velocity of the center of mass of the vortices and $F_c^{\text{sample}}$ is the critical force, both measured for a given disorder realization, and the overline is an average over all disorder realizations for a fixed value of $f$. 
This approach has been used for example in numerical simulations of elastic interfaces in a disordered medium \cite{Bustingorry2008}.\\
\indent We set the relative disorder strength $\alpha_p/\alpha_v \approx 5.10^{-3}$ and we plot in Fig.\ \ref{Fig2} the reduced velocity $v$ with respect to the reduced force $f$ for two different system sizes: $(L_x,L_y)=(100, 50 \sqrt3/2) \lambda_L$ averaged over $N=21$ samples in Fig.\ \ref{Fig2}a, and $(L_x,L_y)=(400, 20 \sqrt3/2) \lambda_L$ averaged over $N=14$ samples in Fig.\ \ref{Fig2}b.

\begin{figure}[!h]
\centering
\includegraphics[width=0.40\textwidth]{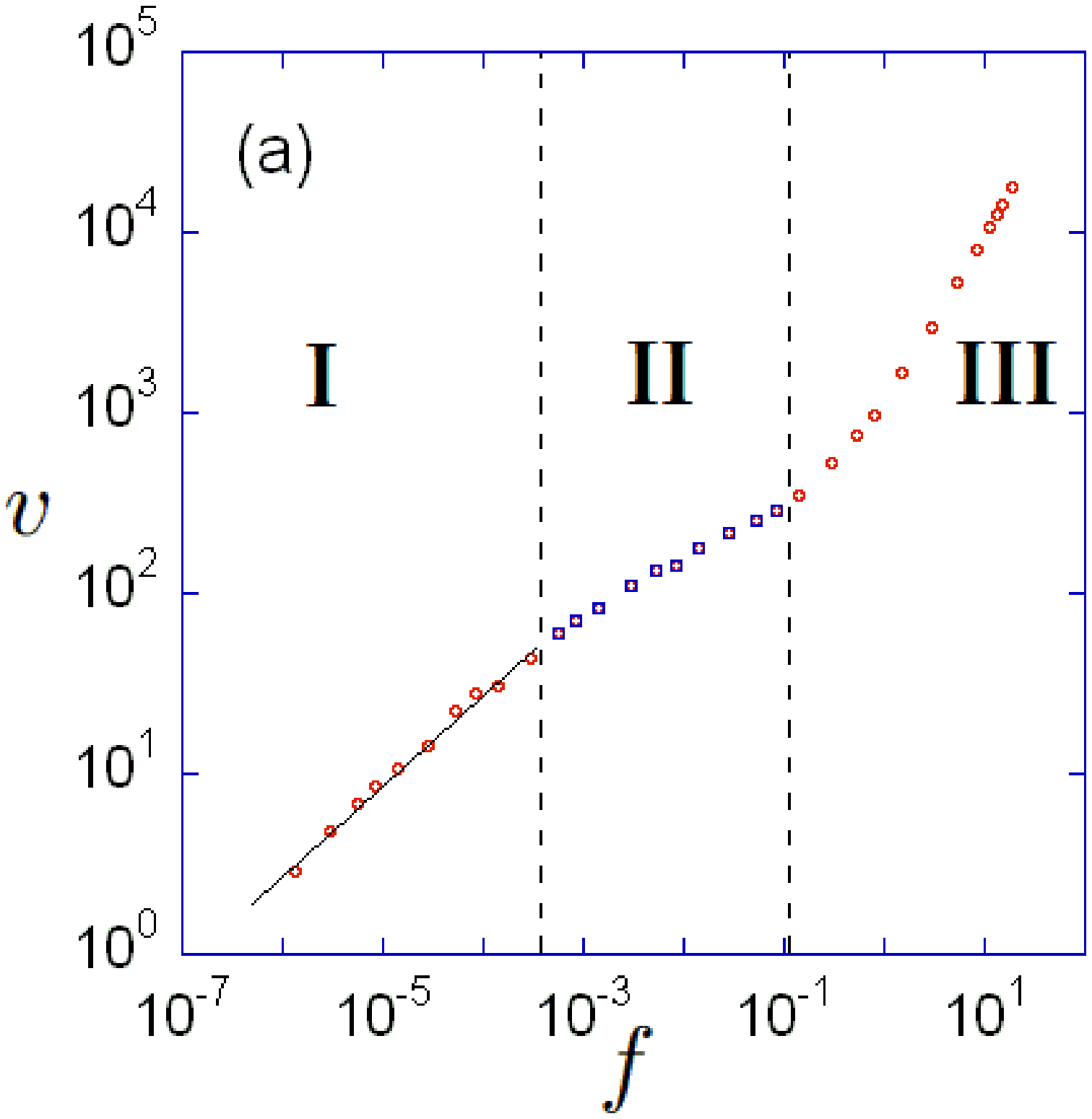}
\includegraphics[width=0.40\textwidth]{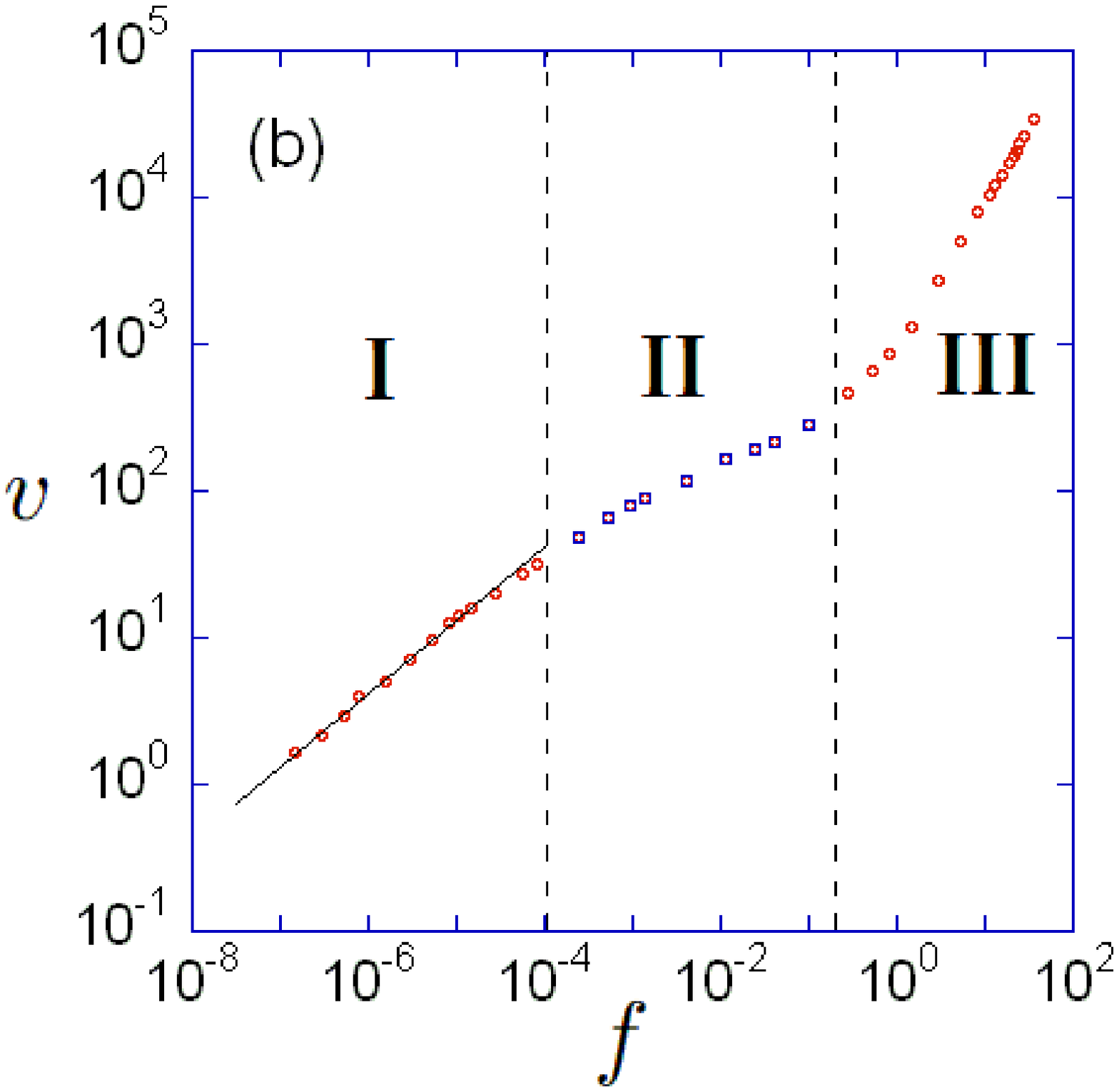}
\caption{\label{Fig2} Plot of the vortex reduced velocity $v$ versus the reduced force $f$ for different system sizes: a) $(L_x,L_y)=(100, 50 \sqrt3/2) \lambda_L$ for $N=21$ samples  b) $(L_x,L_y)=(400, 20 \sqrt3/2) \lambda_L$ for $N=14$ samples. The solid lines indicate a slope of $1/2$ corresponding to the single-particle regime.}
\label{Fig2}
\end{figure}

Three regions appear in Fig.\ \ref{Fig2}. Region I is the manifestation of the finite size effects in the system whose signature is the {\it single-particle regime} \cite{Fily2010}, where $v \sim f^{1/2}$  as shown by the lines of slope $1/2$. Possible hysteretic depinning may be measured in this region for few disorder realizations, with different values $F_c^{up}$ and $F_c^{down}$ of the threshold force when increasing or decreasing the force. However, the width $F_c^{up}-F_c^{down}$ of the hysteresis decreases when the system size increases, which therefore confirms that such hysteresis phenomenon is just a finite size effect. 
In region II a power law regime $v \sim f^{\beta}$ with $\beta < 1$ is measured, which we identify with the critical regime of the continuous depinning transition. Finally, in region III the system is far above the critical depinning threshold and approaches the asymptotic linear behavior $v \sim  f$ obtained in the absence of disorder.

\begin{figure}[!h]
\centering
\includegraphics[width=0.40\textwidth]{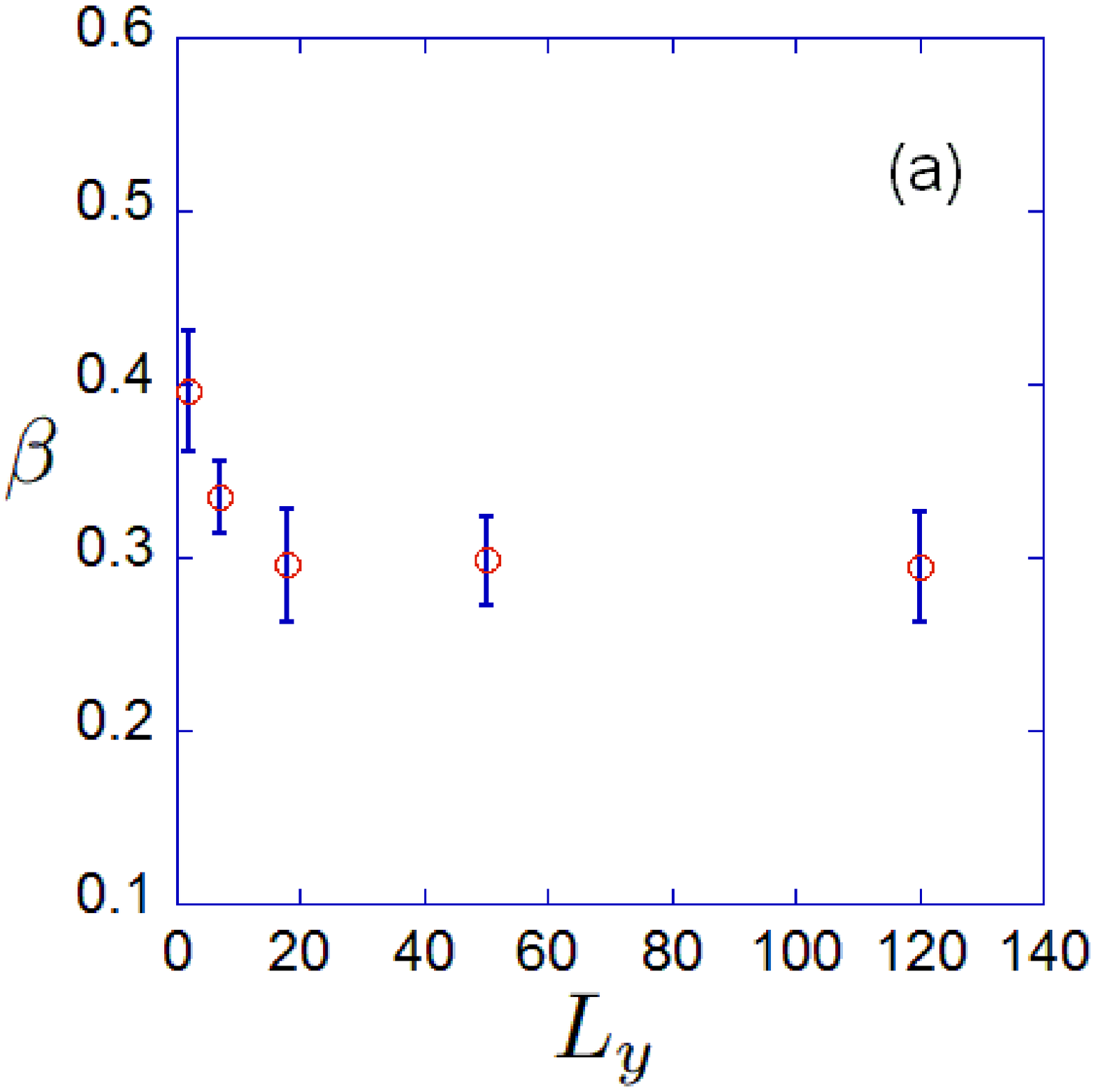}
\includegraphics[width=0.40\textwidth]{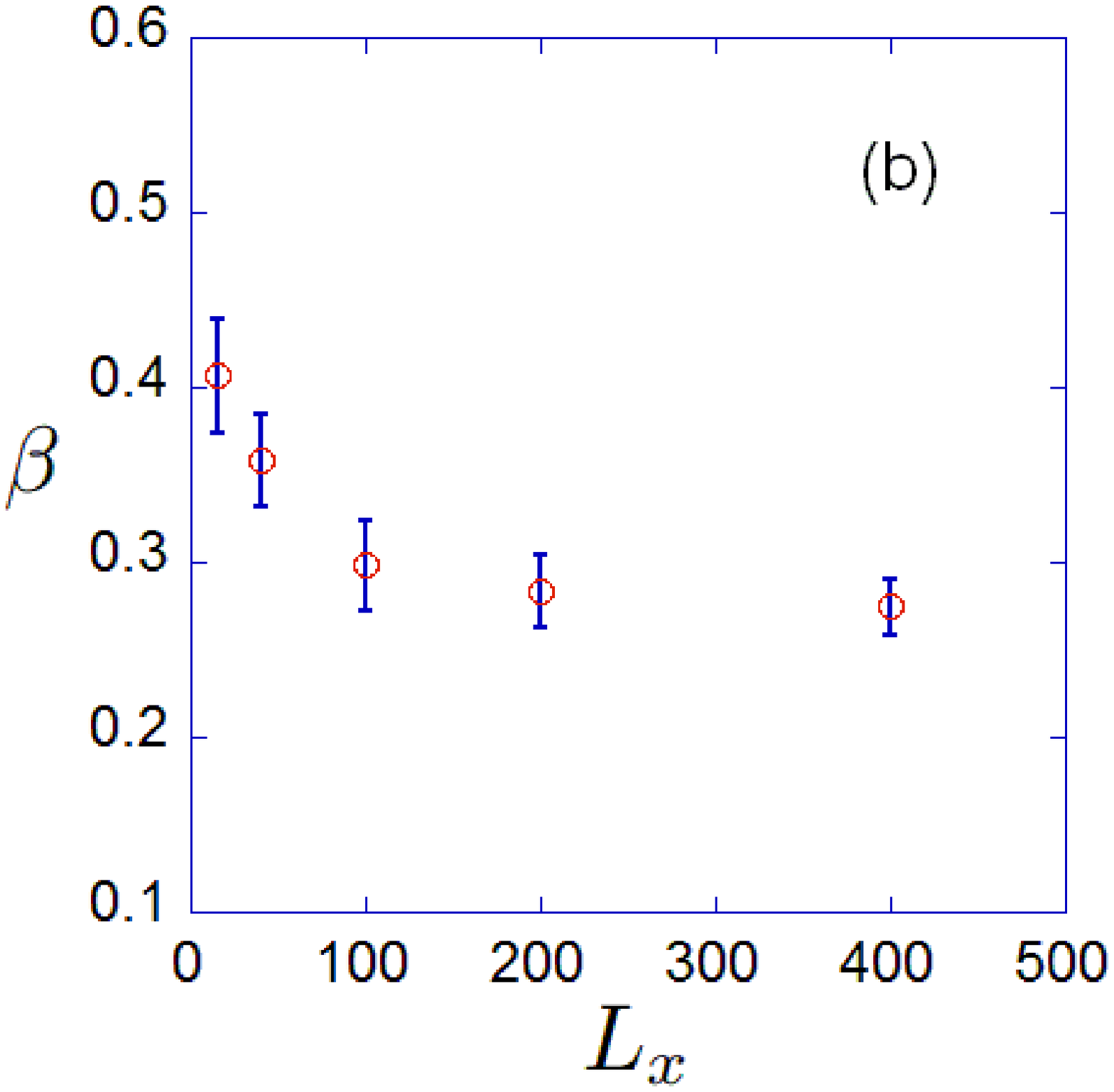}
\caption{\label{Fig3} Depinning exponent $\beta$ extracted from a power law fit in region II of the disorder averaged velocity-force response a) versus $L_y$ for $L_x=100\lambda_L$. Averages are computed over $N=(9, 7, 4, 21, 4)$ samples for respectively $L_y=(2, 7, 18, 50, 120) \sqrt3/2 \lambda_L$ b) versus $L_x$ for various transverse sizes from $L_y=18 \sqrt3/2\lambda_L$ to $L_y=120 \sqrt3/2\lambda_L$. Averages are computed over $N=(44, 47, 39, 12, 27)$ samples for respectively $L_x=(15, 40, 100, 200, 400)\lambda_L$. The error bars correspond to the standard error associated to the value of $\beta$ extracted from the fit.}
\label{Fig3}
\end{figure}

For each system size we compute the depinning exponent $\beta$ from a power law fit in the region II of the disorder averaged $v-f$ curves (see Fig.\ \ref{Fig2}).
Fig.\ \ref {Fig3}a shows the evolution of $\beta$ with respect to the transverse size $L_y$ for a fixed longitudinal size $L_x=100\lambda_L$. We see that the value of $\beta$ for $L_y\geq 18 \sqrt3/2 \lambda_L$ becomes independant of the transverse size $L_y$. In particular $\beta$ does not depend on the basic cell anisotropy since we measure identical values in square basic cells $(L_x,L_y)=(100, 120 \sqrt3/2) \lambda_L$. In Fig.\ \ref {Fig3}b we show the evolution of $\beta$ with respect to the longitudinal size $L_x$ for various transverse sizes $L_y\geq 18 \sqrt3/2 \lambda_L$. It can be seen that $\beta$ has reached a constant value for $L_x \geq 100\lambda_L$. 
Taking the mean value of $\beta$ computed for both $L_x \geq 100\lambda_L$ and $L_y \geq 18 \sqrt3/2 \lambda_L$, we obtain the result $\beta = 0.29\pm 0.03$.\\
Note that taking the mean of individual values $\beta_i$ for each disorder realization, \emph{i.e.} without averaging the velocity over disorder, gives a similar value $\beta = 0.27 \pm 0.04$.

\subsection{Finite-size scaling}
\label{FSS}

Finite-size systems used in numerical simulations lead to rounding and shifting effects in second-order phase transitions that can be analysed within the finite-size scaling theory \cite{Cardy1988}. In addition to providing a clearer picture of the thermodynamic limit of the transition from finite size observations, this approach allows to extract further information about diverging lengths in the system. From a general theorem for disordered systems \cite{Chayes1986} a finite-size scaling length can be defined from the statistical properties of a large number of finite-size samples. The finite-size scaling exponent $\nu_{\mbox{\tiny FS}}$ characterizing the divergence of such a length at the threshold should satisfy the inequality $\nu_{\mbox{\tiny FS}} \geq 2/d$.\\
\indent We first examine the critical force distribution for each system size, and determine their width $\Delta F_c(L_x)$ for the relative disorder strength $\alpha_p/\alpha_v \approx 5.10^{-3}$. 
In Fig.\ \ref{Fig4}a we display 
$\Delta F_c(L_x)$ versus $L_x$ which scales as \cite{Middleton1993a},
$$\Delta F_c(L_x) \sim L_x^{-1/\nu_{\mbox{\tiny FS}}}$$ with $\nu_{\mbox{\tiny FS}}=1.09 \pm 0.07$, in agreement with the inequality $\nu_{\mbox{\tiny FS}} \geq 2/d$.\\

\begin{figure}[!h]
\centering
\includegraphics[width=0.40\textwidth]{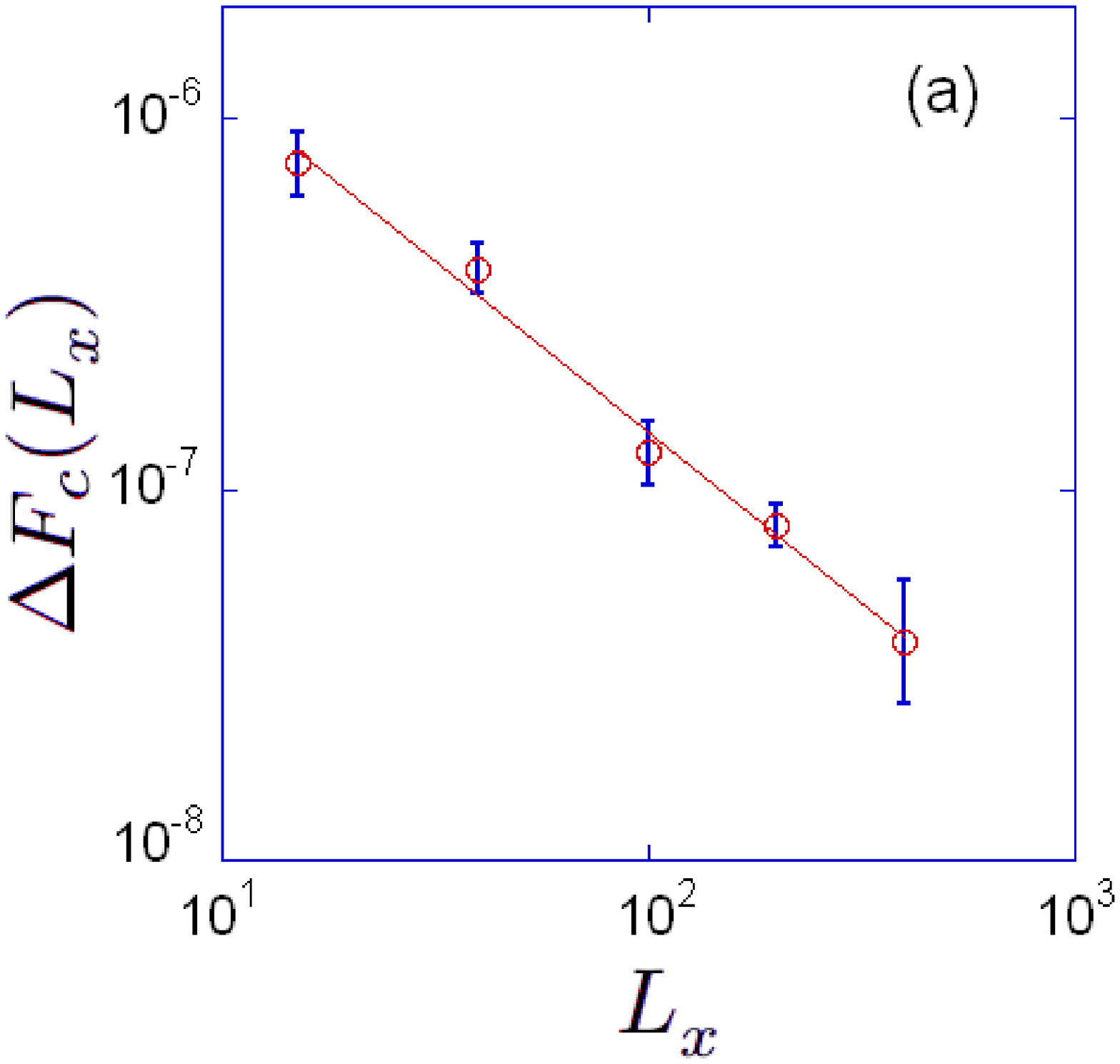}
\includegraphics[width=0.40\textwidth]{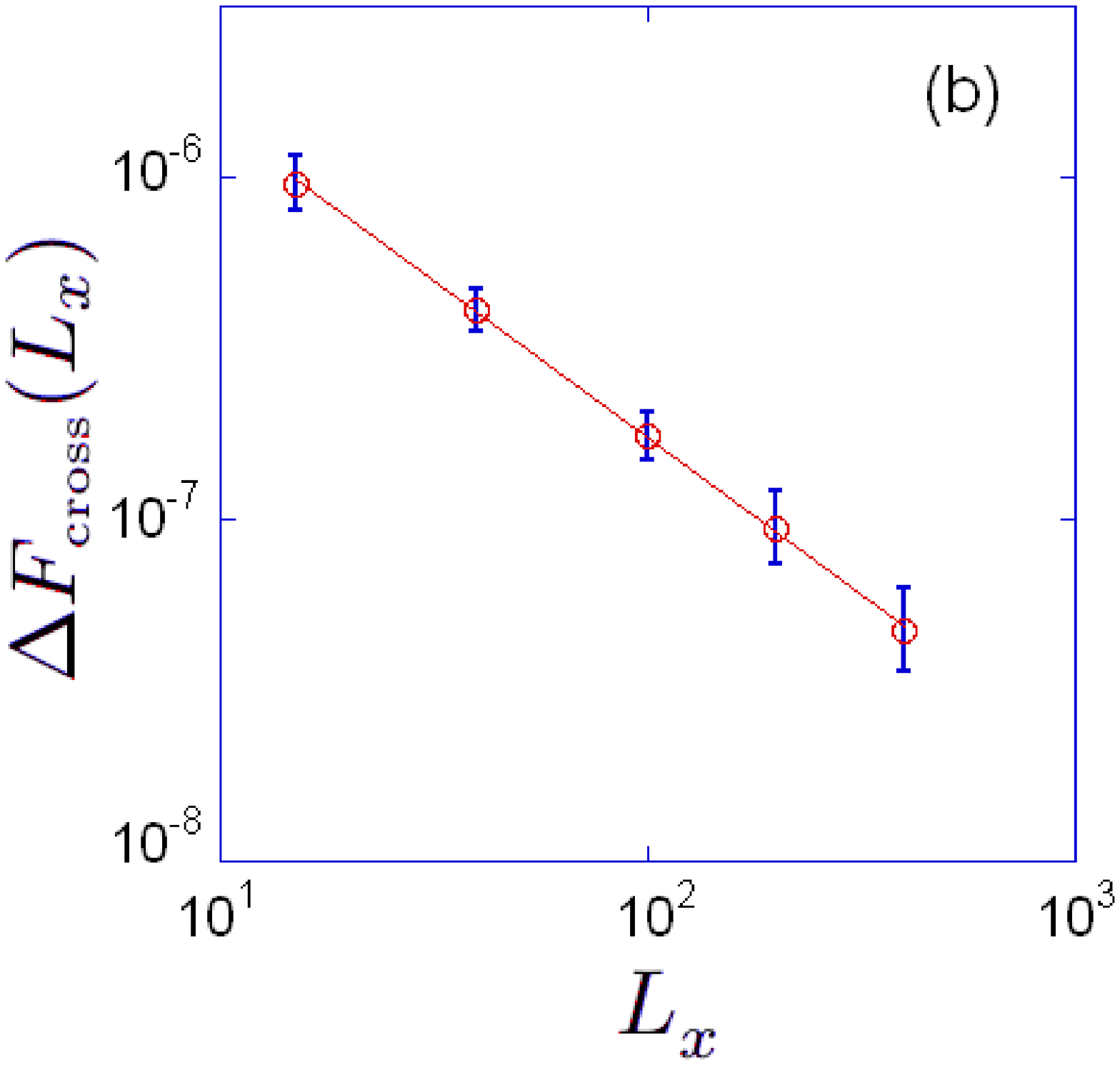}
\caption{\label{Fig4} a) Width of the critical force distribution $\Delta F_c(L_x)$ versus $L_x$. A power-law fit to the data gives the solid line of slope $-1/\nu_{\mbox{\tiny FS}}$ from which we extract $\nu_{\mbox{\tiny FS}}=1.09 \pm 0.07$; b) Width of the crossover force distribution $\Delta F_{\text{cross}}(L_x)$ versus $L_x$. A power-law fit to the data gives the solid line of slope $-1/\nu_{\mbox{\tiny FS}}$ from which we extract $\nu_{\mbox{\tiny FS}}=1.09 \pm 0.02$. The data are the same as those shown in Fig.\ \ref {Fig3}b. The error bars indicate the standard error associated to the value of the width extracted from the distribution.}
\label{Fig4}
\end{figure}

\indent Similarly, for each system size we identify the appearence of finite size effects with the crossover force $F_{\text{cross}}$ between the critical regime and the \emph{single-particle regime} (see regions I and II in Fig.\ \ref{Fig2}). We plot in Fig.\ \ref{Fig4}b the width $\Delta F_{\text{cross}}(L_x)$ of the crossover force distribution, 
which behaves like $$\Delta F_{\text{cross}}(L_x) \sim L_x^{-1/\nu_{\mbox{\tiny FS}}}$$
with the same exponent $\nu_{\mbox{\tiny FS}}=1.09 \pm 0.02$. \\
\indent Furthermore, we expect the correlation length $\xi$ to behave in the critical regime as $\xi \sim (F-F_c^{\infty})^{-\nu}$ where $\nu$ is the intrinsic correlation length exponent, and $F_c^{\infty}$ is the critical force for the infinite system. 
At the crossover force $F_{\text{cross}}$ the correlation length becomes of the order of the system size $L_x$, so that the crossover force $F_{\text{cross}}$ varies with the longitudinal size $L_x$ as
$$F_{\text{cross}} - F_c^{\infty} \sim L_x^{-1/\nu}.$$
Fig.\ \ref{Fig5} shows  $\overline{F_{\text{cross}}} - F_c^{\infty}$ versus $L_x$ where $\overline{F_{\text{cross}}}$ is the disorder averaged crossover force. We determine $F_c^{\infty}$ from the asymptotic behavior of the disorder averaged sample critical force $\overline{F}_{c}$ for large $L_x$ (see the inset of Fig.\ \ref{Fig5}).

\begin{figure}[!h]
\centering
\includegraphics[width=0.47\textwidth]{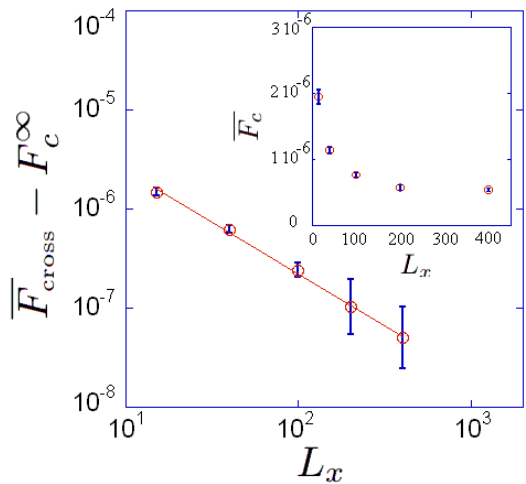}
\caption{\label{Fig5} Plot of $\overline{F_{\text{cross}}} - F_c^{\infty}$  versus $L_x$ for the same data as in Fig.\ \ref{Fig4}. A power-law fit gives $\nu = 1.04 \pm 0.04$. Inset: Averaged sample critical force $\overline{F_c}$ versus $L_x$ from which we extract the asymptotic value $F_c^{\infty}$ for large $L_x$. The error bars represent the standard error of our samples. The large error bars for the two largest system sizes in the main figure come from the statistical errors which become of the order of $\overline{F_{\text{cross}}} - F_c^{\infty}$.}
\label{Fig5}
\end{figure}

From a power law fit shown in Fig.\ \ref{Fig5}, we extract the value $\nu = 1.04 \pm 0.04$ which is consistent with $\nu_{\mbox{\tiny FS}}$ computed previously.
Therefore, a diverging length above the depinning threshold with an exponent $\nu = \nu_{\mbox{\tiny FS}}$ seems to appear. Our results suggest that this length coincides with the intrinsic correlation length $\xi$. We show in Fig.\ \ref{Fig6} the direct computation of the correlation length $\xi$ extracted from the velocity correlation function fit with the functional form $<v_x(0)v_x(x)> \sim C_0 x^{-\kappa} e^{-x/\xi}$ (see \cite{Duemmer2005}). We find that both fit parameters $C_0$ and $\kappa$ are almost constant when $F-F_c^{\infty}$ varies. For a given disorder realization and $(L_x,L_y)=(100, 18 \sqrt3/2) \lambda_L$, Fig.\ \ref{Fig6} shows the evolution of $\xi$ as a function of $F-F_c^{\infty}$.

\begin{figure}[!h]
\centering
\includegraphics[width=0.47\textwidth]{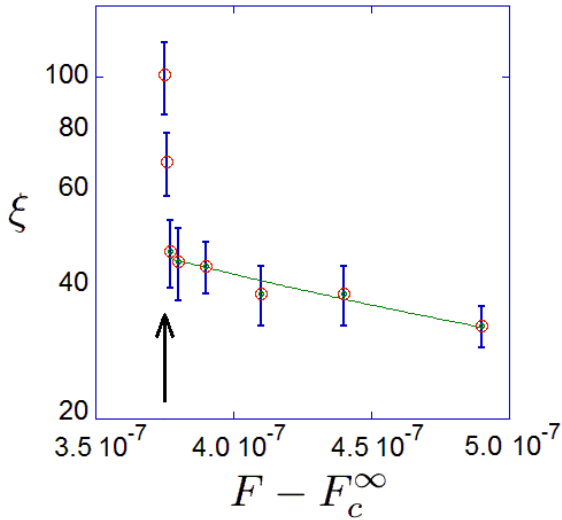}
\caption{\label{Fig6} Plot of $\xi$ versus $F-F_c^{\infty}$ for a given disorder realization and for a system size $(L_x,L_y)=(100, 18 \sqrt3/2) \lambda_L$ at $\alpha_p/\alpha_v \approx 5.10^{-3}$. The error bars indicate the standard error associated to the value of $\xi$ extracted from the fit $<v_x(0)v_x(x)> \sim C_0 x^{-\kappa} e^{-x/\xi}$. The black arrow indicates the value of $F_{\text{cross}} - F_c^{\infty}$.}
\label{Fig6}
\end{figure}

As expected, when $F \rightarrow F_{\text{cross}}^+$ finite size effects actually appear when $\xi \sim L_x$ with $L_x=100\lambda_L$. Furthermore the correlation length $\xi$ diverges as $\xi \sim (F-F_c^{\infty})^{-\nu}$ for $F > F_{\text{cross}}$ in the critical regime, and the power law fit exponent gives $\nu = 1.2 \pm 0.2$ which nicely agrees with Fig.\ \ref{Fig5}. Therefore, our results strongly suggest that, unlike CDWs \cite{Middleton1993a,NarayanFisher1992,*Myers1993a,LeDoussal2002}, the intrinsic correlation length controls both the critical force distribution and the finite-size crossover force.
\\

We can now write a scaling relation between the velocity, the driving force and the system size $L_x$ at $T=0$ as,
$$v \sim L_x^{-\beta/\nu} g(f L_x^{1/\nu})$$
This scaling relation is relevant only close enough to the depinning transition 
and at large enough $L_x$. 

\begin{figure}[!h]
\centering
\includegraphics[width=0.47\textwidth]{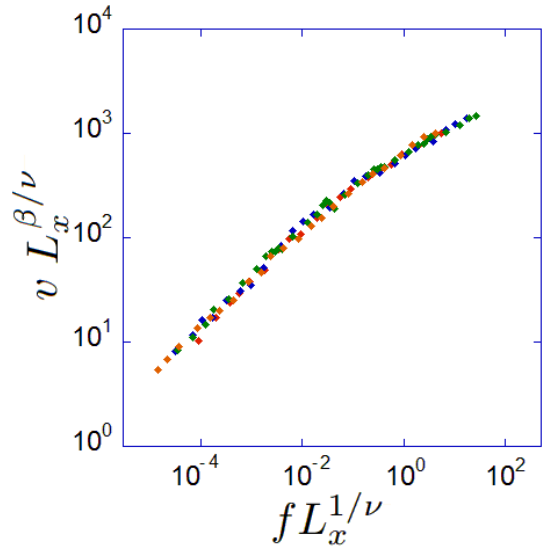}
\caption{\label{Fig7} Plot of $v L_x^{\beta/\nu}$ versus $f L_x^{1/\nu}$ for different system sizes $L_x=(40,100,200,400)\lambda_L$ at $\alpha_p/\alpha_v \approx 5.10^{-3}$. The value of the $\beta$ and $\nu$ exponent is $\beta = 0.29\pm 0.03$ and $\nu=1.04 \pm 0.04$.}
\label{Fig7}
\end{figure}

We take the disorder averaged response $v(f)$ for different system sizes $L_x=(40,100,200,400)\lambda_L$ and various transverse sizes from $L_y=18 \sqrt3/2 \lambda_L$ to $L_y=120 \sqrt3/2 \lambda_L$, and for the relative disorder strength $\alpha_p/\alpha_v \approx 5.10^{-3}$. In Fig.\ \ref{Fig7} we plot $v L_x^{\beta/\nu}$ versus $f L_x^{1/\nu}$ using the value of $\beta = 0.29\pm 0.03$ and $\nu=1.04 \pm 0.04$, retaining only the points of the region I and II. A good collapse of the data onto a single curve is obtained.
 The behavior of the scaling function $g(x)$ is $g(x)\sim x^{\beta}$ for $x \gg 1$, and in contrast to ordinary phase transitions $g(x) \rightarrow 0$ for $x \rightarrow 0$. This unusual property has already been observed in numerical simulations of the depinning transition of driven interfaces \cite{Nowak2008}. It is due to the \emph{single-particle regime} which can partially hides usual finite-size effects.\\

\section{Finite temperatures}
\label{T!=0}

\subsection{Critical exponent $\delta$}
\label{delta}

In this section we study the finite temperature effects on the depinning transition. For $F<F_c$ at $T>0$ the vortex mean velocity does not vanish since thermal fluctuations provide sufficient energy to overcome local energy barriers: this results in a rounded depinning transition. Using the analogy with standard critical phenomena, a power law response $v_{F=F_c} \sim T^{1/\delta}$ can be defined at $F=F_c$.
We show in Fig.\ \ref{Fig8} the typical reduced velocity-temperature response that we obtain.
Two different behaviors can be seen in the $v(T)$ response. Below the critical force, $v$ goes to zero faster than a power law resulting in concave $v(T)$ curves, while above the critical force $v$ approches a nonzero limit as $T$ goes to zero resulting in convex $v(T)$ curves with an horizontal asymptote on the left. The effective critical force $\overline{F_c^*}$ is defined as the force at which the convexity changes. In agreement with a second-order phase transition, we can extrapolate at $\overline{F_c^*}$ a power law response $v_{F=\overline{F_c^*}} \sim T^{1/\delta}$ (dashed line in Fig.\ \ref{Fig8}) from which we measure the critical exponent $\delta^{-1}=0.28 \pm 0.05$, where the error bar is deduced from the different lines one can draw to extrapolate the power-law behavior.\\
\indent We would like to emphasize that this new determination of the disorder averaged critical force is consistent with the value of $\overline{F_c}$ analysed at $T=0$ for the same samples in section \ref{T=0}. Going back to the $v(f)$ curves at $T=0$ and using the effective critical force $\overline{F_c^*}$, we find $\beta=0.28 \pm 0.05$ which nicely agrees with the value $\beta=0.29 \pm 0.03$ extracted from the zero temperature simulations of section \ref{T=0}. This larger uncertainty comes from the smaller number of samples simulated at finite temperatures.\\
\indent Finally, note that we observe an identical mean value $\delta^{-1} = 0.28$ when taking the mean of individual values $\delta^{-1}_i$ obtained from the velocity-temperature curves for each disorder realization.

\begin{figure}[!h]
\centering
\includegraphics[width=0.47\textwidth]{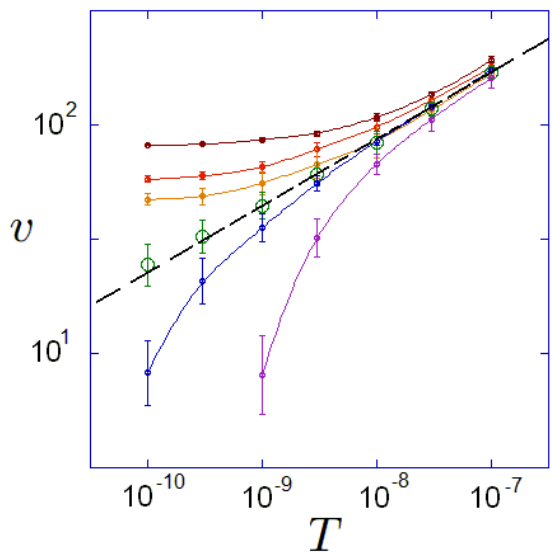}
\caption{\label{Fig8} Reduced velocity $v$ versus temperature $T$, averaged over $N=6$ samples for fixed values of $f = (10, 4, 3, 0, -2, -7) \times 10^{-4}$. The system size is $(L_x,L_y)=(400,20 \sqrt3/2)\lambda_L$ and the relative disorder strength is $\alpha_p/\alpha_v \approx 5.10^{-3}$. The dashed line is the extrapolation of $v(T)$ at the effective critical force $\overline{F_c^*}$ from which we extract the value of $\delta^{-1}$. The error bars correspond to the standard error.}
\label{Fig8}
\end{figure}

\subsection{Different pinning strengths - Scaling law}
\label{scaling}

For different system sizes with $L_x \geq 100\lambda_L$ and $L_y \geq 18 \sqrt3/2 \lambda_L$, we investigate several values of the relative pinning strength $\alpha_p/\alpha_v \approx (2, 3, 5, 7)\times 10^{-3}$. The number of samples is $N=1$ for $\alpha_p/\alpha_v \approx 2.10^{-3}$ and $L_x=1000\lambda_L$, $N=2$ for $\alpha_p/\alpha_v \approx 3.10^{-3}$ and $L_x=400\lambda_L$, $N=8$ for $\alpha_p/\alpha_v \approx 5.10^{-3}$ and $L_x=(100,400)\lambda_L$, $N=4$ for $\alpha_p/\alpha_v \approx 7.10^{-3}$ and $L_x=400\lambda_L$.  
For each pinning strength, we compute both the critical exponents $\beta$ and $\delta^{-1}$ and we display their evolution respectively in Fig.\ \ref{Fig9}a and Fig.\ \ref{Fig9}b. The values of $\beta$ and $\delta^{-1}$ are found to be independent of the relative pinning strength.\\

\begin{figure}[!h]
\centering
\includegraphics[width=0.40\textwidth]{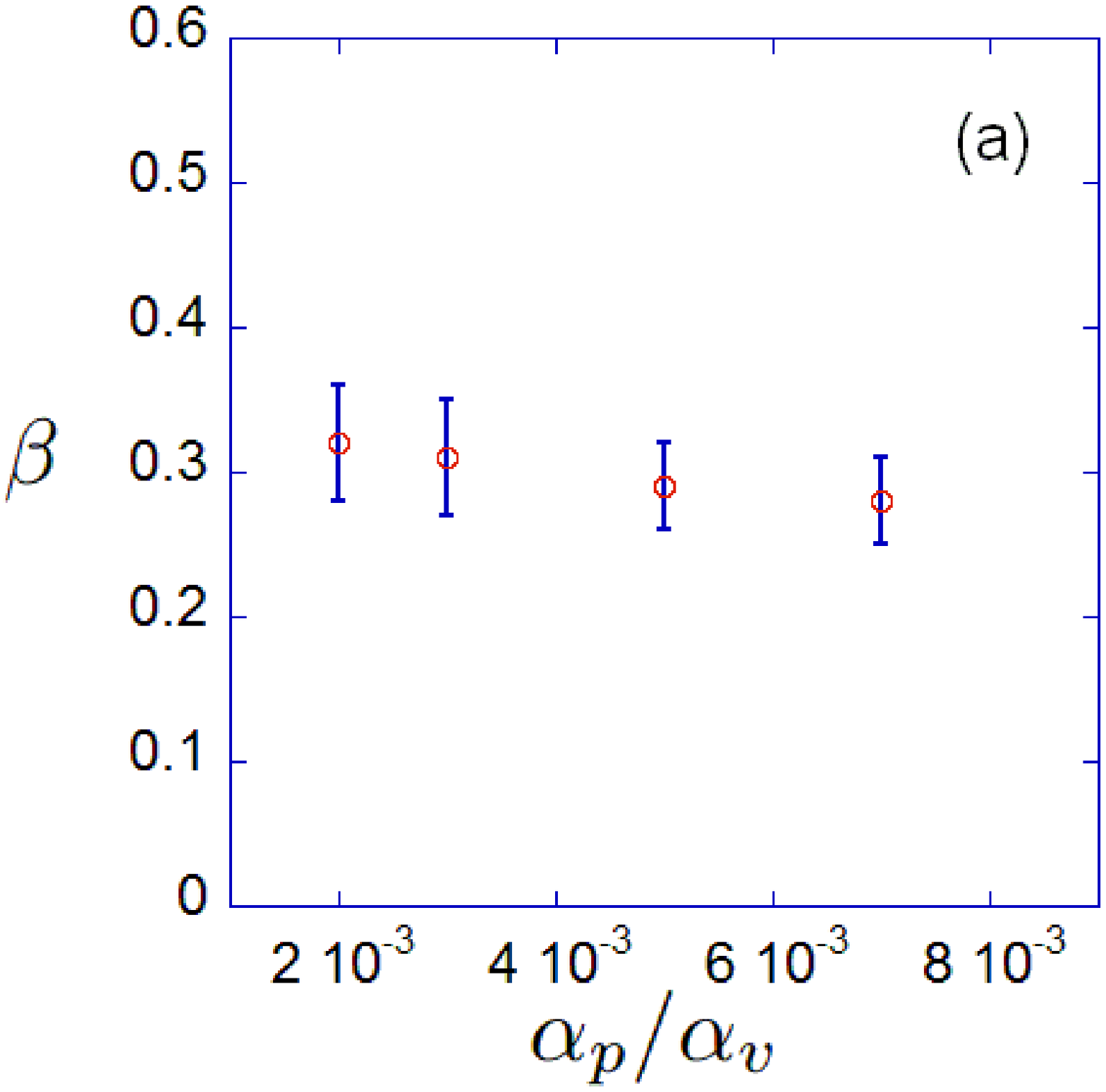}
\includegraphics[width=0.40\textwidth]{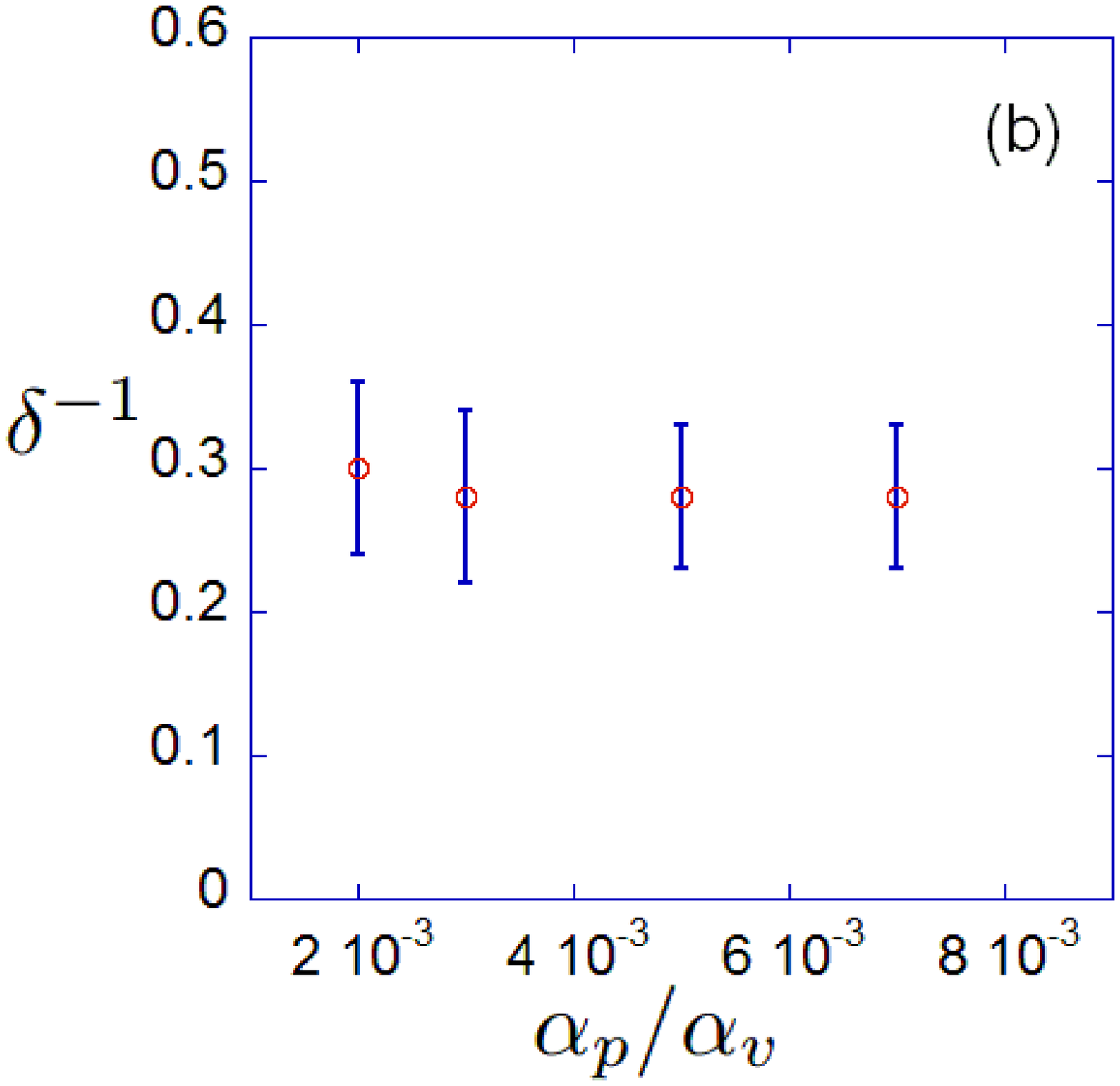}
\caption{\label{Fig9} Evolution with the relative pinning strength $\alpha_p/\alpha_v$ of the critical exponent a) $\beta$ and b) $\delta^{-1}$.}
\label{Fig9}
\end{figure}

Because of the power law responses of the velocity with $T$ and $f$, we assume that the velocity $v$ is a generalized homogenous function of driving force $f$ and temperature $T$. We want to express the relation between $v$, $f$, and $T$ in terms of dimensionless quantities and to make this relation independent of the prefactors in the two power law responses. We define $v_0$ and $T_0$ as
$$v_{f>0,T=0}=v_0 f^{\beta}, v_{f=0}=v_0 \left(\frac{T}{T_0}\right)^{1/\delta}$$
and we introduce the dimensionless velocity $\widetilde{v}=v/v_0$ and the dimensionless temperature $\widetilde{T}=T/T_0$. We can then define the following scaling ansatz,
$$\widetilde{v}|f|^{-\beta} = S_{\pm}(\widetilde{T}|f|^{-\beta \delta})$$
where $S(x)$ is the scaling function with two branches $S_{+}$ and $S_{-}$ corresponding to $f>0$ and $f<0$, respectively. The power law dependencies $\widetilde{v}_{f>0,T=0}=f^{\beta}$ and $\widetilde{v}_{f=0}=\widetilde{T}^{1/\delta}$ imply that $S(x)$ has the following asymptotical behavior:
$$\lim_{x \rightarrow 0} S_{+}(x) = 1\ \ \text{and}\ \lim_{x \rightarrow \infty} x^{-1/\delta}S_{\pm}(x) = 1$$
The $S_{+}$ branch goes asymptotically to the horizontal axis for $\widetilde{T} \rightarrow 0$ defining a driving dominated regime, while the $S_{\pm}$ branches have both an oblique asymptote with slope $\delta^{-1}$ for $f \rightarrow 0$ defining a temperature dominated regime.
The change of variables $(v,T) \rightarrow (\widetilde{v},\widetilde{T})$ is equivalent to choosing the intersection of the asymptotes as the origin of coordinates.\\

\begin{figure}[!h]
\centering
\includegraphics[width=0.49\textwidth]{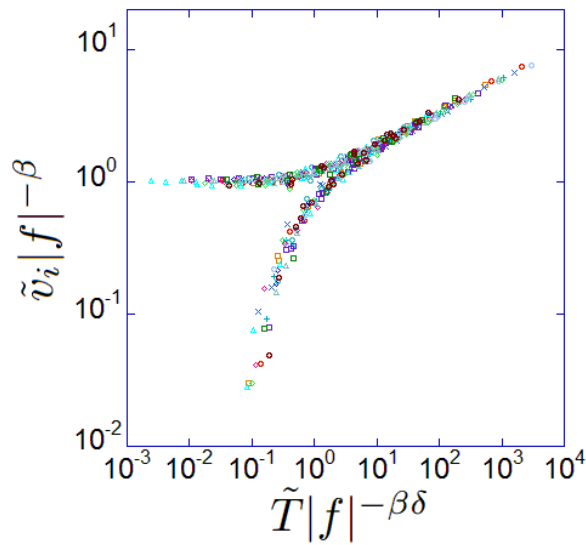}
\caption{\label{Fig10} Scaling plot of $\widetilde{v}_i|f|^{-\beta}$ versus $\widetilde{T}|f|^{-\beta \delta}$ with $\beta=0.29 \pm 0.03$ and $\delta^{-1}=0.28 \pm 0.05$ for $N=14$ samples with different pinning strengths $10^3 \times \alpha_p/\alpha_v \approx (2, 3, 5, 7$), system sizes $L_x=(100,400,1000)\lambda_L$ and disorder realizations. Each sample value of $\beta$ and $\delta^{-1}$ is allowed to vary within the error bars.}
\label{Fig10}
\end{figure}

To show the existence of this scaling behavior, we compute the velocity without averaging over the disorder, \emph{i.e.} we compute $\widetilde{v}_i$ for each of the $N=14$ samples that we have studied, with $i=1,..,N$. This choice allows to better see if all data scale on a single curve. Different system sizes $L_x=(100,400,1000)\lambda_L$ \emph{i.e.} $N_v=(5000,8000,20000)$, different pinning strengths $10^{3} \times \alpha_p/\alpha_v \approx (2, 3, 5, 7)$ and different realizations of the random position of the pinning centers have been investigated. In Fig.\ \ref{Fig10} we plot $\widetilde{v}_i|f|^{-\beta}$ with respect to $\widetilde{T}|f|^{-\beta \delta}$ and a good collapse of all individual samples to a single curve with two branches is obtained. The values of $F_c$ used to compute $f$ are sample specific. 
The two prefactors $v_0$ and $T_0$ are not universal and their value varies for each sample, but the scaling functions $S_{\pm}$ remain unchanged. We verified that averaging over disorder as previously defined, we obtain the same scaling functions.\\
\indent The collapse of our data implies that 
a true critical regime is observed with power law responses for the velocity with respect to the driving force and to the temperature. Moreover the critical depinning exponent $\beta$ and the thermal exponent $\delta^{-1}$ do not depend on the strength and the positions of the pins. Theses results suggest some degree of universality of our model for the elastic depinning of vortices.

\section{Discussion}
\label{discuss}

In our study we find a continuous elastic depinning transition for weak pinning strength. Both the depinning exponent $\beta$ and the thermal exponent $\delta$, and the scaling function linking velocity $v$, temperature $T$ and driving force $f$ show some degree of universality with respect to disorder. 
A large variety of $\beta$ values has been found in the literature as cited in section \ref{intro}. Our value is close to the value $\beta \approx 1/3$ measured for interfaces \cite{Duemmer2005} in a space dimension $d=2$. However, our result should rather be compared to other periodic systems.
For example, our result is close to the value $\beta=0.35$ recently measured in anisotropic periodic systems (stripe systems) \cite{OlsonReichhardt2011} for which the displacement field dimension is $N=2$ in a $d=2$ dimensional space. 
Much studies appear in the literature for the case $N=1$, in particular for CDWs where numerical simulations \cite{Myers1993} in $d=2$ give $\beta=0.65\pm 0.05$ and $\nu=0.5 \pm 0.1$, whereas our results give very different values $\beta=0.29\pm 0.03$ and $\nu=1.04 \pm 0.04$. Also, we find $\nu \approx 1$ for the intrinsic correlation legnth exponent while $\nu=1/2$ is expected for CDWs above threshold. Moreover, our results of section \ref{FSS} suggest $\nu=\nu_{FS}$ in contradiction with CDWs. Actually, the poor agreement between our results and CDWs may originate from the fact that $N=1$ for CDWs whereas $N=2$ in our vortex simulation.\\
We note that inserting $\nu=1$ in the scaling relation $\nu=1/(2-\zeta)$ which originates from the statistical tilt symmetry \cite{Narayan1993} gives $\zeta=1$. When inserting both values in the hyperscaling relation $\beta=\nu (z-\zeta)$ \cite{Nattermann1992,Narayan1993,LeDoussal2002} (where $z$ and $\zeta$ are respectively the dynamic and roughness critical exponents) together with $z=5/4$ \cite{Majumdar1992}, we obtain $\beta=1/4$ which is close to our result $\beta=0.29\pm 0.03$ but may not be satisfactory since $\zeta=0$ is expected to hold in the hyperscaling relation.

Finally, going back to the connection between the depinning transition and absorbing-state phase transitions, we note that our result $\beta=0.29\pm 0.03$ is close to the value found in $d=1$ directed percolation or conserved directed percolation, and that $\nu= 1.04 \pm 0.04$ is close to the value of $\nu_{\perp}$ in $d=1$ directed percolation.

\section{Conclusion}
\label{conclu}

In this paper we have numerically studied the 2D superconductor vortex dynamics in random media. A crossover from plastic dynamics dominated by disorder to elastic dynamics dominated by elasticity is found. We investigated the depinning transition in the elastic regime. Above the depinning threshold all the vortices depin together and flow in elastically coupled rough static channels. 
 Three dynamical regimes are found: the first one is controlled by the finite-size of the simulation box leading to the so-called \emph{single-particle regime}, the second one is the critical regime from which a power law response $v \sim f^{\beta}$ at $T=0$ can be extracted with $\beta=0.29 \pm 0.03$, while the third one is the recovery of the ohmic regime. A finite temperature analysis gives the thermal rounding exponent $\delta^{-1}=0.28 \pm 0.05$ defined by the power law response $v \sim T^{1/\delta}$ at the threshold force. A finite-size scaling analysis at zero temperature suggests (in contrast with CDWs) the existence of a unique diverging length above the depinning threshold with an exponent $\nu= 1.04 \pm 0.04$, which controls the critical force distribution, the finite-size crossover force distribution and the intrinsic correlation length. A scaling law for the temperature and the force dependence of the velocity is found, consistent with a second order phase transition.
Both critical exponents and the scaling function are independent of the disorder (strength and position of the pins) in the range of parameters we studied, indicating some degree of universality. Nevertheless the comparison with other similar periodic systems may suggest that large universality can not be found for 2D elastic depinning.
We expect that such elastic depinning transition occurs in relatively weak pinning regimes of superconductor vortices. Within transport measurements where the voltage response to an applied current is equivalent to the velocity response to an applied force, the critical exponents $\beta$ and $\delta$ may be accessible from finite temperature measurements at different driving currents.

\section*{Acknowledgement}

We gratefully acknowledge the generous grant of computer time from the Orl\'eans-Tours {\it CaSciModOT Cluster} at the \emph{Centre de Calcul Scientifique de la R\'egion Centre} - France. We also wish to thank Pierre Le Doussal, Kay Wiese and Grégory Schehr for helpful discussions.

\end{document}